\documentclass[12pt,aps,onecolumn,nobibnotes]{revtex4}
\usepackage{graphicx}%
\usepackage{dcolumn}
\usepackage{amsmath}
\usepackage{amssymb}
\font\sqi=cmssq8
\def\DR{\rm I\kern-1.45pt\rm R}
\def\DC{\kern2pt {\hbox{\sqi I}}\kern-4.2pt\rm C}
\def\DH{\rm I\kern-1.5pt\rm H\kern-1.5pt\rm I}
\newcommand{\bs}{\mbox{\boldmath $\sigma$}}
\begin{document}
\begin{abstract}
\noindent
We show  that the oscillators on a sphere and pseudosphere
are related,   by the so-called Bohlin transformation,
with the Coulomb systems on the pseudosphere:
the even states of an oscillator yields
 the conventional Coulomb system on pseudosphere,
 while the odd states yield the
Coulomb system on pseudosphere in the presence of magnetic flux tube
 generating half spin.
In the higher dimensions the  oscillator and Coulomb(-like) systems
are connected in the similar way.
In particular, applying the Kustaanheimo-Stiefel transformation to the
oscillators on sphere and pseudosphere,
we obtained the  preudospherical generalization of MIC-Kepler problem
describing  three-dimensional  charge-dyon system. 
\end{abstract}
\title[Short Title]{How to relate  the oscillator and Coulomb systems\\
 on spheres and pseudospheres?}
\author{{Armen Nersessian}}
\thanks{On leave of absence from  the Joint Institute for Nuclear Research,
141980 Dubna, Russia  
 and Yerevan State University,
 A. Manoogian St., 1, Yerevan, 375025 Armenia}
\affiliation{The Abdus Salam Internationaal Centre for Theoretical Physics,
Trieste, Italy}
\maketitle

\section{Introduction}
The ($d-$dimensional) oscillator and Coulomb systems
are most known representatives
of mechanical systems possessing  hidden symmetries which define the
 $su(d)$ symmetry algebra for the oscillator, and $so(d+1)$
for the Coulomb system.
The hidden symmetry has  a very transparent meaning
in the case of oscillator,
 while in  the case of the Coulomb system  it has
a more complicated interpretation
 in terms of geodesic flows of a $d$-dimensional sphere
.
On the other hand, the transformation $r=R^2$ converts
 the $(p+1)-$dimensional radial Coulomb problem
in  $2p-$dimensional
 radial oscillator one, both in classical
and quantum cases, where the $r$ and $R$
denote the radial 
coordinates of Coulomb and oscillator systems,
 respectively (see, e.g. \cite{ta}).
In three distinguished cases, $p=1,2,4$,
 one can establish the complete  correspondence
 between the Coulomb and the oscillator systems,
 by  using  the so-called
Bohlin (or Levi-Civita) \cite{bohlin}, Kustaanheimo-Stiefel \cite{ks}
and Hurwitz \cite{h} transformations, respectively.
 This transformations  assume
 the reduction of the oscillator system by the action of $Z_2$,
$U(1)$, $SU(2)$ groups, respectively, and yield the
Coulomb-like systems specified by the presence
of  monopoles \cite{ntt,mic,su2}.
On the other hand, the  oscillator and
Coulomb systems admit the  generalizations to a $d-$di\-men\-si\-onal
sphere and two-sheet hyperboloid (pseudosphere) with radius $R_0$
 given by the potentials \cite{sphere,sphere1}
\begin{equation}
V_{osc}=\frac{\alpha^2 R_0^2}{2}\frac{{\bf x}^2}{{x}^2_{d+1}},\quad V_{C}=
-\frac{\gamma}{R_0}\frac{x_{d+1}}{|{\bf x}|},
\end{equation}
where ${\bf x}, x_{d+1}$ are the (pseudo)Euclidean coordinates of ambient
space $\DR^{d+1}$($\DR^{d.1}$):
$\epsilon{\bf x}^2+ x^2_{d+1}=R_0^2,\;\;
 \epsilon=\pm 1$. The  $\epsilon =1$
corresponds to the sphere, and $\epsilon =-1$ to the pseudosphere.
 These systems possess nonlinear hidden symmetries providing
them  with the properties similar to those
 of conventional oscillator and Coulomb systems and
have been investigated  from  many viewpoints 
(see,  e. g. \cite{2o} and refs therein).
 
{\it How  to relate the oscillator and Coulomb systems on the
spheres and pseudospheres?} \\
Recently this problem was considered in   Refs.\cite{kmp},
 where   the oscillator
and Coulomb systems on spheres   
were related by some complicated  mappings containing the transitions 
to imaginary coordinates. The geometrical origin of these mapping was not
clarified there, as well as the reductions to the Coulomb-like systems with
the monopoles, and the relations of the motion constants responsible for
hidden symmetries, were  not considered there.
In our recent paper  with G.Pogosyan \cite{np}  we  established 
 the transparent correspondence  between oscillator 
and Coulomb systems on (pseudo)spheres
for the simplest, two-dimensional, case ($p=1$).
We have shown that, in the stereographic projection, the conventional
Bohlin transformation relates the   two-dimensional oscillator
on the (pseudo)sphere with the  Coulomb systems on pseudosphere,
as well as those interacting with specific external magnetic fields.
This simple  construction allows immediately  connect the
motion constants defining   the hidden symmetry of the 
systems under consideration, as well as to clarify the mappings 
suggested in  \cite{kmp}.This construction   can be straightly used for  
higher- dimensional cases ($p=2,4$),  subject to obtain the 
the pseudospherical analogs of  the known Coulomb-like systems, 
specified by the presence of monopoles: the so-called MIC-Kepler \cite{z,mic} 
and  $SU(2)$ Kepler\cite{su2} problems.
In present  paper we give a   detailed description of this
construction  for the $p=1$ case  corresponding to  the Bohlin
transformation ({\it Section 2}), for the $p=2$ case which corresponds to
the Kustaanheimo-Stiefel one ({\it Section 3}) and discuss the $p=4$ case
corresponding to Hurwitz transformation ({\it Section 4}).
  
\section{The Bohlin transformation}
Let us introduce  the complex coordinate
$z$ parameterizing the sphere
by the complex projective plane $\DC P^1$ and
the two-sheeted hyperboloid by  the Poincar\'e  disks ${\cal L}$:
\begin{equation}
{\bf x}\equiv x_1+ix_2=R_0\frac{2z}{1+\epsilon z\bar z},\quad
 x_3=R_0\frac{1-\epsilon {z\bar z}}{1+\epsilon {z\bar z}}.
\label{x}\end{equation}
In these terms  the metric takes the K\"ahler form
\begin{equation}
 ds^2=R_0^2\frac{4dz d\bar z}{(1+\epsilon{z\bar z})^2},
\label{met}\end{equation}
while $R_0x_k$ define the isometries
of the Kahler structure
($su(2)$ if $\epsilon=1$ and $su(1.1)$ if $\epsilon=-1$).
The lower hemisphere and the lower sheet of the hyperboloid are
parametrized by the unit disk $|z|<1$, while the upper hemisphere
and the upper sheet of hyperboloid, by its outside, and transform
into each other  by the inversion $z\to 1/z$.
Since in the $R_0\to \infty$ limit the lower
 hemisphere (the lower sheet of hyperboloid) converts into the
 whole two-dimensional plane,
for the correspondence with conventional
oscillator and Coulomb
problems, we have to restrict ourselves by those defined
on the lower hemisphere and the lower sheet of hyperboloid (pseudosphere).

Let us equip the  oscillator's phase space
$T^*\DC P^1$ ($T^*{\cal L})$ by the symplectic structure
\begin{equation}
\omega=d\pi\wedge dz+ d{\bar\pi}\wedge d{\bar z}
\label{ss}\end{equation}
and  the rotation generators (defining  $su(2)$ algebra if $\epsilon=1$
and $su(1.1)$  if $\epsilon=-1$)
\begin{equation}
{\bf J}\equiv \frac{iJ_1 -J_2}{2}=\pi +\epsilon{\bar z}^2\bar\pi,
\;\; J\equiv \frac{\epsilon J_3}{2}=i(z\pi-{\bar z}{\bar\pi}).
\label{j}\end{equation}
In these terms,
the oscillator's Hamiltonian is given
 by the expression
\begin{equation}
{ H}^{\epsilon}_{osc}(\pi, {\bar\pi}, z, {\bar z})=
\frac{{\bf J}{\bf\bar J}+\epsilon J^2}{2R_0^2}+
\frac{\alpha^2 R_0^2}{2}\frac{{\bf x}^2}{{x}^2_{3}}=
\frac{(1+\epsilon z{\bar z})^2\pi{\bar\pi}}{2R_0^2}
+\frac{2\alpha^2R_0^2{z\bar z}}{(1-\epsilon{z\bar z})^2}.
\label{ho}\end{equation}
The hidden symmetry is given by the
complex (or vectorial) constant of motion \cite{sphere1}
\begin{equation}
{\bf I}=I_1+iI_2=\frac{{\bf J}^2}{2R_0^2} +
\frac{\alpha^2R_0^2}{2}\frac{{\bf\bar  x}^2}{x^2_3},
\label{I}\end{equation}
which defines, together  with $J$ and $H_{osc}$ ,  the cubic algebra
\begin{equation}
 \{{\bf I}, J\}=2i{\bf I},\;\;\{{\bf\bar I},{\bf I}\}=4i
\left(\alpha^2 J +\frac{\epsilon JH_{osc}}{R_0^2}-\frac{J^3}{2R_0^4}\right).
\label{Ia}\end{equation}
The energy surface of the oscillator on the (pseudo)sphere
$H^\epsilon_{osc}=E$  reads
\begin{equation}
\frac{\left(1-(z\bar z)^2\right)^2{\pi}{\bar\pi}}{2R_0^4}+
2\left(\alpha^2+\epsilon \frac{E}{R^2_0}\right) z{\bar z}
=\frac{E}{R^2_0}\left(1+(z\bar z)^2\right).
\label{os}\end{equation}
Now, performing the canonical Bohlin transformation \cite{bohlin}
\begin{equation}
w=z^2,\quad p=\frac{\pi}{2z},
\label{boh}\end{equation}
we convert the energy surface of the oscillator (\ref{os})
onto the one of the Coulomb system on the pseudosphere:
\begin{equation}
\frac{(1-w\bar w)^2{p}{\bar p}}{2r_0^2}-
\frac{\gamma}{r_0}\frac{1+w{\bar w}}{2|w|}={\cal E}_{C},
\label{C}\end{equation}
where
\begin{equation}
r_0=R_0^2,\quad  \gamma=\frac{E}{2},
\quad -2{\cal E}_{C}=\alpha^2+\epsilon\frac{E}{r_0}.
\label{gam}\end{equation}
The constants of motion of the oscillators, $J$ and ${\bf I}$
(which   are  equal
 on the energy surfaces (\ref{os})) converted, respectively
 into the doubled
angular momentum and the doubled Runge-Lenz vector
 of the Coulomb system
\begin{equation}
J\to 2J_{C},\quad
 {\bf I}\to 2{\bf A},\quad {\bf A}=
-\frac{iJ_{C}{\bf J}_{C}}{r_0}+
{\gamma}\frac{{\bf\bar x}_{C}}{|{\bf x}_{C}|},
\end{equation}
where ${\bf J}_{C}$, $J_{C}$, ${\bf x}_{C}$ denote the rotation generators
and  the pseudo-Euclidean coordinates  of the Coulomb system.

It is easy to obtain from (\ref{Ia}) the symmetry algebra of the reduced system
\begin{equation}
 \{{\bf A}, J\}=i{\bf A},\;\;\{{\bf\bar A},{\bf A}\}=-4i
\left({\cal H}_{C}+\frac{J^2_C}{r_0^2}\right)J_C.
\label{Aa}\end{equation}
{\it Hence, the Bohlin transformation of the
classical isotropic oscillator on the (pseudo)sphere
yields the  classical Coulomb problem on the pseudosphere.}

The  quantum-mechanical  counterpart of the energy surface
(\ref{os}) is the Schr\"odinger equation
 \begin{equation}
{\cal H}^{\epsilon}_{osc}(\alpha, R_0| \pi,{\bar\pi},z,{\bar z})\Psi(z,{\bar z})=
E\Psi(z,{\bar z}),
\label{1}\end{equation}
with the quantum Hamiltonian defined
(due to the two-dimensional origin of the system) by the expression
(\ref{ho}), where  $\pi,\bar\pi$  are the momenta operators
(hereafter we assume $\hbar=1$)
\begin{equation}
\pi=-i\frac{\partial}{\partial{z}},\quad
{\bar\pi}=-i\frac{\partial}{\partial{\bar z}} .
\end{equation}
The energy spectrum of this system
is given by the expression (see e.g. \cite{2o} and refs therein)
\begin{equation}
E= {\tilde\alpha}(N+1)+\epsilon\frac{(N+1)^2}{2R^2_0},
\quad N=2n_r+|M|,\quad n_r=0,1,\ldots\quad.
\label{oenergy}\end{equation}
where ${\tilde\alpha}=\sqrt{\alpha^2+1/(4R^4_0)}$,
$M$ is the eigenvalue of $J$, $N$ is the principal quantum number,
 $n_r$ is the radial quantum number,
\begin{equation}
 |M|,\;N=1,\ldots,
N_{max}=\left\{
\begin{array}{cc}
\infty\;, &{\rm if }\; \epsilon=1
\cr
[2{\tilde\alpha}R^2_0]-1\;,&
{\rm if }\; \epsilon=-1
\end{array}
\right.
\end{equation}
So,  the the number of levels in the energy spectrum of the
oscillator is infinite on the sphere and finite on the pseudosphere.

The quantum-mechanical correspondence
between oscillator and Coulomb systems is  more complicated,
because the  Bohlin transformation (\ref{boh}) maps
 the $z$-plane into the
two-sheeted Riemann surface, since ${\rm arg}\;w\in[0,4\pi)$.
 Thus, we have to supply the quantum-mechanical  Bohlin  transformation
with the reduction by the $Z_2$ group action,
 choosing either
 even  ($\sigma=0$) or odd ($\sigma=1/2$) wave functions
 \begin{equation}
  \Psi_\sigma (z, \bar z) =
\psi_\sigma (z^2, {\bar z}^2)\left(\frac{z}{\bar z}\right)^{2\sigma}:\quad
\psi_{\sigma}(|w|, {\rm arg} w + 2\pi)=
\psi_{\sigma}(|w|, {\rm arg} w ).
\label{3}\end{equation}
 This implies that the range  of
definition of $w$ can be restricted, without loss of generality,
 to ${\rm arg}\; w\in[0,2\pi)$.
In that case, the resulting system is the Coulomb
problem on the hyperboloid given by the Schr\"odinger equation
\begin{equation}
H^{-}_{C}(\gamma, r|p_{\sigma}, {\bar p}_\sigma,w,\bar w)
\psi_\sigma={\cal E}_C\psi_{\sigma}
\end{equation}
where   $\gamma, {\cal E}_C, r$ are given by (\ref{gam}),
 and the
momenta operators are of the form
\begin{equation}
{ p}_\sigma = -i\frac{\partial}{\partial w}-\frac{\sigma}{iw},
\quad{\bar p}_\sigma = -i\frac{\partial}{\partial\bar w}
+\frac{\sigma}{i{\bar w}}.\end{equation}
Hence, the resulting Coulomb system includes the interaction
 with the   magnetic vortex (an infinitely thin solenoid)
with the magnetic flux $\pi\sigma $  and  zero strength
 $rot{\sigma}/{w}=0$. Such a composites are typical representatives
of the anyonic systems with the spin  $\sigma$.
{\it So, we get  a conventional} $2d$ {\it Coulomb problem
on the hyperboloid at } $\sigma=0$ {\it
  and those with half spin  generated by the
magnetic flux, at} $\sigma=1/2${\it .}
Taking into account the relations (\ref{gam}),
one can rewrite  the oscillator's energy spectrum (\ref{oenergy})
as follows
\begin{equation}
\sqrt{\frac{1}{4r_0^2}-\epsilon\frac{2\gamma}{r_0}-2{\cal E}_C}=
\frac{2\gamma}{N+1}-\epsilon\frac{N+1}{2r_0}.
\label{inter}\end{equation}
 From this expression one can easily
obtain the   energy spectrum of the reduced system on the pseudosphere
\begin{equation}
{\cal E}_C=-\frac{N_\sigma(N_\sigma+1)}{2r^2_0}-
\frac{\gamma^2}{2(N_\sigma+1/2)^2},
\end{equation}
where
\begin{equation}
    N_\sigma=n_r+m_\sigma,\quad m_\sigma=M/2, \quad
n_r, m_\sigma-\sigma, N_\sigma-\sigma=0,1,\ldots, N_\sigma^{max}-\sigma.
\end{equation}
Here $m_\sigma$  denotes the eigenvalue of the angular momentum of
the reduced system, and $n_r$ is the radial quantum number of the
initial (and reduced) system.
Notice, that the magnetic vortex shifts the energy levels
 of the two-dimensional Coulomb system
 which is nothing else than  the reflection of
Aharonov-Bohm effect.\\
It is seen, that the whole spectrum of the oscillator on pseudosphere
($\epsilon =-1$) transforms in the spectra of the constructed
Coulomb systems on the pseudosphere, while for the oscillator on
the sphere ($\epsilon =1$) the positivity of
l. h. s. of (\ref{inter})  restrict the admissible values of $N_\sigma$.
 So,  only the part of the spectrum of the oscillator
on the sphere transforms into the spectrum of Coulomb system.
Hence, in  both  cases we get the same  result
 \begin{equation}
N^{max}_\sigma  =\left[\sqrt{r_0\gamma}-(1/2+\sigma)\right].
\end{equation}
\section{Kustaanheimo-Stiefel transformation}
It is easy to see that the $2p-$dimensional oscillator on (pseudo)sphere
can be  connected with the $(p+1)-$dimensional
Coulomb-like systems on pseudosphere   likewise in the higher
dimensions ($p=2,4$).
Indeed, in stereographic coordinates, the
oscillator on $2p$-dimensional
(pseudo)sphere  is described  by the
Hamiltonian  system given by (\ref{ss}), (\ref{ho}), where
 the following replacement is performed
$(z,\;\pi)\to (z^a,\;\pi_a)$, $a=1,\ldots, p$
with the summation over these indices.
Consequently,  the oscillator's energy surfaces  are
of the form (\ref{os}).
Further reduction to the $(p+1)$-dimensional Coulomb-like
system on pseudosphere must be 
similarly followed  in the  corresponding 
reduction in the flat case \cite{mic,su2}. 
Since  $|{\bf u}|=z{\bar z}$ in all three cases, we  can  interpret  ${\bf u}$ 
 as the stereographic coordinates of the reduced system,
consequently interpreting the last one as the 
Coulomb-like system on $(p+1)$-dimensional pseudosphere.

For example, if $p=2$, we should reduce  the four-dimensional
oscillator by the Hamiltonian action of $U(1)$ group given by the generator
$$J=i(z\pi-{\bar z}{\bar\pi}).$$
For this purpose, we  have to  fix the level surface
\begin{equation}
J=2s
\label{Js}\end{equation}
and factorize it by the $U(1)$- Hamiltonian flow, choosing six
   $U(1)$-invariant  stereographic coordinates
in the form of  conventional
Kustaanheimo-Stiefel transformation \cite{ks,mic}
\begin{equation}
{\bf u}=z{\bs}{\bar z},\quad
{\bf p}=\frac{z{\bs}\pi+{\bar \pi}{\bs}{\bar z}}{2({z\bar z})},
\label{ks}\end{equation}
where $\bs$ are Pauli matrices. 

As a result,
the reduced
symplectic structure reads
\begin{equation}
d{\bf p}\wedge d{\bf u} +
s\frac{({\bf u}\times d{\bf u})\wedge d{\bf u}}{|{\bf u}|^3},
\label{ss2}\end{equation}
the oscillator's energy surface
takes the form
\begin{equation}
\frac{(1- {{\bf u}}^2)^2}{8r_0^2}({\bf p}^2 +\frac{s^2}{{\bf u}^2})-
\frac{\gamma}{r_0}\frac{1+{\bf u}^2}{2|{\bf u}|}={\cal E}_C,
\label{C3}\end{equation}
where ${\bf u}$ denote the stereographic coordinates 
of three-dimensional pseudosphere, while $r_0$, $\gamma$, ${\cal E}_C$
are defined  by the expressions (\ref{gam}).

So, we get the energy surface of the  pseudospherical analog
of a Coulomb-like system describing
the interaction of two non-relativistic dyons, which was proposed in 
\cite{z} and is known as the MIC-Kepler system.

In the coordinates of ambient space the  potential  of pseudospherical 
MIC-Kepler system looks as follows
\begin{equation}
V_{MIC}=\frac{s^2}{r^2_0}\left(\frac{x^2_4}{2|{\bf x}|^2}-
2\right)-\frac{\gamma}{r_0}\frac{x_4}{|{\bf x}|}
\end{equation}

To quantize the system, we should replace 
the equations (\ref{os}),(\ref{Js})
by the following spectral problem
\begin{equation}
{\hat{\cal H}}_{osc}(\pi,{\bar\pi}, z,{\bar z})\Psi(z,{\bar
z})=E_{osc}\Psi(z,{\bar z}),\quad
{\hat J}_0(\pi,{\bar\pi}, z,{\bar z})\Psi(z,{\bar z}) =2s\Psi(z,{\bar z})
\label{sp}\end{equation}
where the momenta $\pi_\alpha,\bar\pi_\alpha$ are replaced by the operators
\begin{equation}
\pi_\alpha=-i\frac{\partial}{\partial z^\alpha},
\quad{\bar \pi}_\alpha=-i\frac{\partial}{\partial{\bar  z}^\alpha},
\label{piq}\end{equation}
and the appropriate ordering in the Hamiltonian is assumed.

The second equation in the (\ref{sp}) can be resolved by the substitution 
of the anzats
\begin{equation}
\Psi_s(z,{\bar z})=\psi_s({\bf u}){\rm e}^{is\lambda}
:\quad 
[{\hat J}_0\lambda]=i,
\quad \lambda =is\log\frac{z^1}{{\bar z}^1},
\label{psis}\end{equation}
which reduces the first equation in (\ref{sp})(i.e., the
 oscillator's Schr\"odinger equation)
to those corresponding to the generalized MIC-Kepler system  (\ref{C3}),
where
\begin{equation}
{\hat {\bf p}}_s={\rm e}^{-is\lambda}{\hat {\bf p}}{\rm e}^{is\lambda}
=-i\frac{\partial}{\partial {\bf u}}-s{\bf A}({\bf u}), 
\end{equation}
with ${\bf  A}({\bf u})$ being the vector potential of Dirac's monopole with 
 singularity directed along axes $u_3$, and ${\hat{\bf p}}$
be defined by the second expression in  (\ref{ks}), where
$\pi$,${\bar\pi}$ are
given by  operators (\ref{piq}) placed at right.
 
The requirement that  $\Psi$ to be single-valued wave function,
leads  $s$ to be integer or half-integer, i.e. the Dirac's
quantization condition.

Solving the  Schr\"odinger equation , one gets 
the oscillator's energy spectrum  \begin{equation}
E= {\tilde\alpha}(N+2)+\epsilon\frac{(N+2)^2-2}{2R^2_0},
\quad N=2n_r+|L|,
\label{oenergy2}\end{equation}
where ${\tilde\alpha}=\sqrt{\alpha^2+1/(4R^4_0)}$,
$L$ is the eigenvalue of complete angular momentum,
 $N$ is the principal quantum number,
 $n_r$ is the radial quantum number,
\begin{equation}
\begin{array}{c}
2|s|=0,1,\ldots, L, \quad
|L|, N=1,\ldots,N_{max}\\
N_{max}=\left\{
\begin{array}{cc}
\infty\;, &{\rm if }\; \epsilon=1
\cr
\left[{\tilde\alpha}R^2_0\left(1+\sqrt{1+2/({\tilde\alpha}R_0^2)^2}\right)\right]-2\;,&
{\rm if }\; \epsilon=-1
\end{array}
\right.
\end{array}
\end{equation}
Completely similar  to the previous 
case, we can get from this expression  
 the energy spectrum of the MIC-Kepler system 
on pseudosphere:
\begin{equation}
{\cal E}_C=-\frac{(n_r+ |L_s|)(n_r+|L_s|+2)}{2r^2_0}-
\frac{\gamma^2}{2(n_r+|L_s|+1)^2},
\label{micen}\end{equation}
where
\begin{equation}
l_s=L/2 , \quad |l_s|, n_r+|l_s|=|s|,|s|+1,\ldots,N_s^{max}.
\end{equation}
Here $l_s$  denotes the eigenvalue of the complete angular momentum of
the reduced system, and $n_r$ is the radial quantum number of the
initial (and reduced) system. Similar to the $p=1$ case, one should get,
that 
$$N^{max}_s+1=\left[\sqrt{r_0\gamma -\frac{1}{2r^2_0}}\right].$$ 
It is convenient to introduce the new quantum number
$$k\equiv n_r+|L_s|-|s|,\quad k=0,1,\ldots, N_s^{max}-|s|,$$
and re-write the  expression (\ref{micen}) as follows
\begin{equation}
{\cal E}_C=-\frac{(k+|s|)(k+|s|+2)}{2r^2_0}-
\frac{\gamma^2}{2(k+|s|+1)^2}.
\label{micen1}\end{equation}
It is seen, that the degenerancy of the reduced system   the same,
 as in the usual MIC-Kepler problem \cite{mic}, viz $k(k+|s|-1)$,.

It is  pleasure to notice, that  the spherical 
generalization of MIC-Kepler system has also been  presented  on  the Colloquium,
 which was constructed by V.Gritsev, Yu.  Kurochkin and V. Otchik \cite{kurochkin}.

\section{Discussion: The Hurwitz transformation}
We have shown, that applying the standard Bohlin/ Kustaanheimo-Stiefel
transformations to the stereographic (conformal-flat) coordinates of the 
two-/ four-dimensional oscillators on sphere and
pseudosphere yield the pseudospherical  two-dimensional Coulomb and 
(three-dimensional) MIC-Kepler systems,  respectively. It is  obvious, from
above-presented consideration, that  the relation of eight-dimensional
oscillator on (pseudo)sphere and of the pseudospherical analog of the
so-called $SU(2)$ Kepler (or Yang-Coulomb ) system \cite{su2} would 
be completely similar to the mentioned cases.

For establishing such a connection (and constructing the pseudospherical
$SU(2)$ Kepler system) we should perform the Hamiltonian reduction of the 
eight-dimensional oscillator by the  $SU(2)$ group action
action    
\begin{equation}
z^a\to z^a g, \;\; g{\bar g}=1,\quad g\in{\DH},\; z^a\in \DH^2
\label{body}\end{equation}
where  $z^1, z^2$ are quaternions, parameterizing stereographic coordinates
of eight dimensional (pseudo)sphere.
The spatial stereographic coordinates of the 
reduced system should be chosen in the form of standard Hurwitz
transformation \cite{h,su2}
\begin{equation}
u=2z_1{\bar z}_2,\quad u_5=z_1{\bar z}_1- z_2{\bar z}_2,\quad\quad
u\in{\DH},\; u_5\in {\DR}
\end{equation}
and completed  with the conjugated momenta and isospinning coordinates as
well.
The potential of the pseudospherical SU(2) Kepler system would be of the
form similar to the MIC-Kepler one,
 \begin{equation}
V_{SU(2)-Kepler}=\frac{j(j+1)}{r_0^2}\left(\frac{x^2_6}{2{\bf x}^2}-2\right)-
\frac{\gamma}{r_0}\frac{x_6}{2|{\bf x}|}
\label{C31}\end{equation}
where $({\bf x}, x_6)$   denote the  coordinates 
the ambient space of  five-dimensional pseudosphere, $j(j+1)$ is the
eigenvalue of
operator ${\cal J}_i^2$ defining the $SU(2)$ group action (\ref{body}),
 while $r_0$, $\gamma$,
are given  by the expressions (\ref{gam}).
The kinetic term  of the Hamiltonian would include the interaction with the
vector potential of five-dimensional  $SU(2)$
monopole \cite{yang}.

\section{ Acknowledgments.}I am grateful to George Pogosyan for suggesting
this  problem, fruitful discussions
and  kind invitation to participate on the XXIII Colloquium on
Group Theoretical Methods in Physics. 
I would like to thank J. Daboul, D. Fursaev, I. V. Komarov, Yu. A. Kurochkin, 
U.  Sukhatme and V. M. Ter-Antonyan  for useful comments and  interest in  work,
and A. D. Ozer for carefully reading and editing the manuscript.

The financial support of the  International Center of Theoretical 
Physics in Trieste, where this work was completed, is also acknowledged.

\end{document}